\documentclass[
 reprint,
superscriptaddress,
 amsmath,amssymb,
 aps,
prl, 
]{revtex4-2}

\usepackage{graphicx}
\usepackage{dcolumn}
\usepackage{bm}
\usepackage{hyperref}

\begin{document}


\title{Tuning the fundamental periodicity of the current-phase relation in multiterminal diffusive Josephson junctions}

\author{Venkat Chandrasekhar}
 \affiliation{Department of Physics and Astronomy, Northwestern University, 2145 Sheridan Road, Evanston, IL 60208, USA}
\email{v-chandrasekhar@northwestern.edu}

\date{\today}

\begin{abstract}
Conventional superconductor/insulator/superconductor (SIS) Josephson junctions, devices where two superconductors are separated by a tunnel barrier are technologically important as elements in quantum circuits, particularly with their key role in superconducting qubits. 
 An important characteristic of Josephson junctions is the relation between the supercurrent $I_s$ and the phase difference $\varphi$ between them.  For SIS junctions, the current-phase relation is sinusoidal and $2\pi$ periodic.  Other types of Josephson junctions, where the material between the superconductors is a weak link or a normal metal (N) may have non-sinusoidal current-phase relations that are still $2\pi$ periodic.  We show here that a multi-terminal diffusive SNS Josephson junction with 4 superconducting contacts  can show a current phase relation between two of the contacts that is a superposition of $2\pi$ and $4\pi$ periodic components whose relative strength is controlled by the phase difference between the other two contacts, becoming $2\pi$ or $4\pi$ periodic for certain values of this phase difference. This tunability might have applications in tailoring the Hamiltonians of superconducting quantum circuits.        
\end{abstract}

\maketitle

The relation between the supercurrent $I_s$ and the phase difference $\varphi$ is one of the defining characteristics of Josephson junctions.  In a conventional superconductor-insulator-superconductor (SIS), this relation is typically sinusoidal, $I_s = I_c \sin \varphi$, where $I_c$ is the critical current of the junction, with a fundamental periodicity of $2\pi$ \cite{tinkham2004introduction}.   The sinusoidal, nonlinear, current-phase relation has found many applications in superconducting circuits \cite{vanDuzer} and superconducting qubits \cite{SuperconductingQubitReview}.  For superconducting qubits, the sinusoidal current phase relation defines the cosine Josephson potential of the quantum Hamiltonian \cite{tinkham2004introduction}. 
 The SIS junction, however, is only one example of a Josephson junction.   Superconducting devices with weak links such as superconducting microbridges and superconductor/normal-metal/superconductor (SNS) junctions frequently have non-sinusoidal current-phase relations \cite{golubov_current-phase_2004}.  In particular, SNS junctions where the normal metal N has a non-trivial topological band structure have been reported to have non-sinusoidal current-phase relations with a $4\pi$ periodicity \cite{toplogical4pi} whose experimental signatures include modified Fraunhofer diffraction patterns \cite{TI-fraunhofer1} and missing Shapiro steps \cite{wiedenmann_4-periodic_2016, jabani-topo-shapiro} in topological proximity effect devices.  SNS junctions where the relative strengths of 2$\pi$ and 4$\pi$ components of the current-phase relation can be tuned have also been discussed recently in the context of the Josephson diode effect \cite{PhysRevB.108.214520, akhmerov-diodes-2023}.  Here we discuss a SNS device with a diffusive normal metal and 4 superconducting contacts.  The current-phase relation between two of the superconducting contacts can be modulated as a function of the phase difference between the other two superconducting contacts, going from a $4\pi$ periodic current-phase relation when the phase difference is 0 to a $2\pi$ periodic current-phase relation when the phase difference is $\pi$.  At intermediate values of the phase difference, the current-phase relation is a sum of $2\pi$ and $4\pi$ contributions whose relative strength can be controlled by the phase.  Similar results have been reported earlier where the tunability of the current-phase relation was found to be a function of the transparency of the normal-metal/superconductor interfaces \cite{volkov_prb}.  Here we show that the effect can be controlled by varying the lengths of the normal metal wires, which is more amenable to experimental verification as reproducibly fixing or even determining the interface transparency is difficult. 
 The tunability of the current phase relation may have applications in tailoring the Hamiltonians of superconducting qubits.  
 \begin{figure}
\includegraphics[width=6cm]{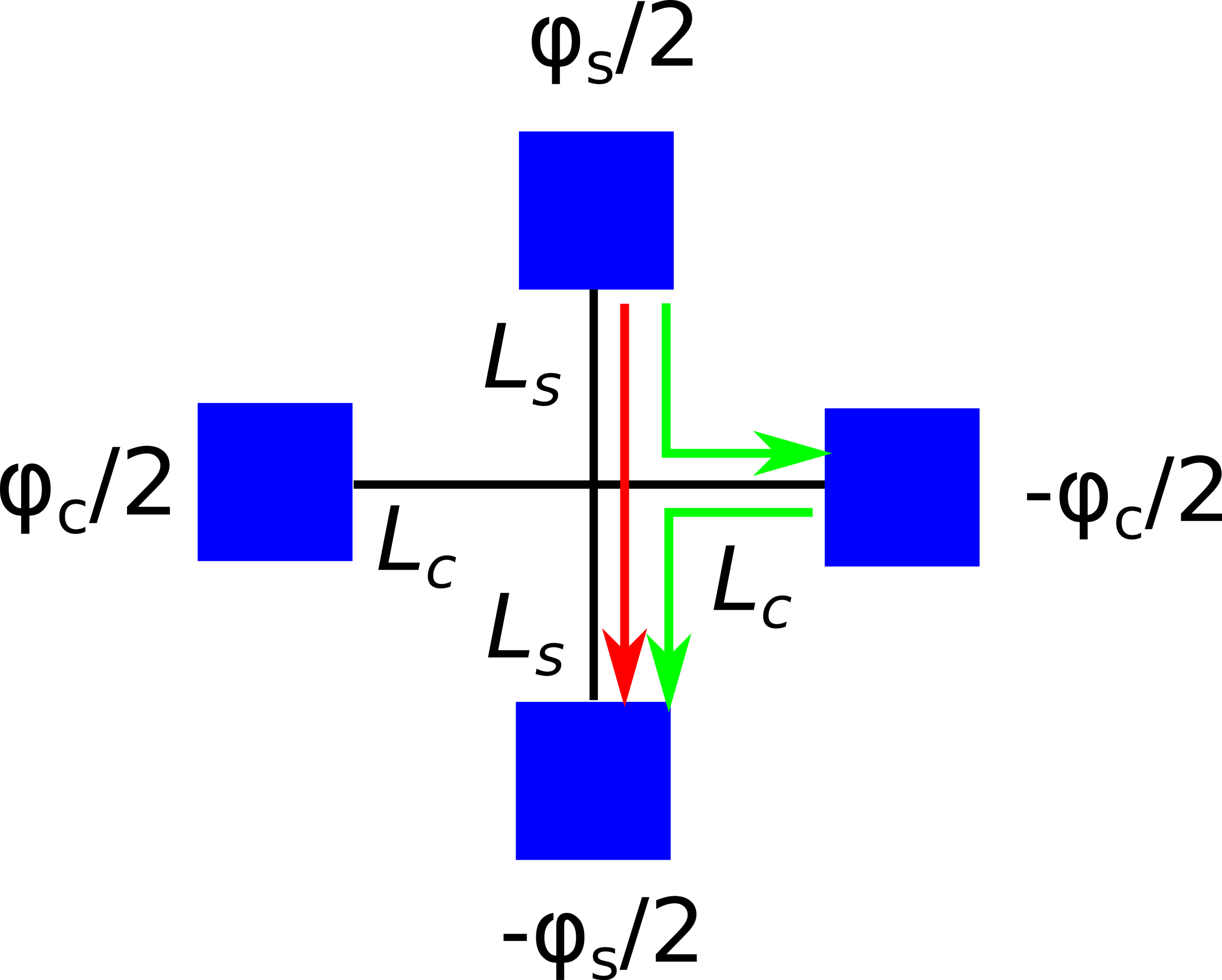}
\caption{Schematic of geometry of the device.  Blue rectangles correspond to superconducting reservoirs with phases imposed as noted, while black lines correspond to one-dimensional normal metal wires with lengths as noted.  Red and green arrows denote some of the paths that supercurrents can take between the superconducting reservoirs.}
\label{fig:Fig1}
\end{figure}
\begin{figure*}
\includegraphics[width=17.5cm]{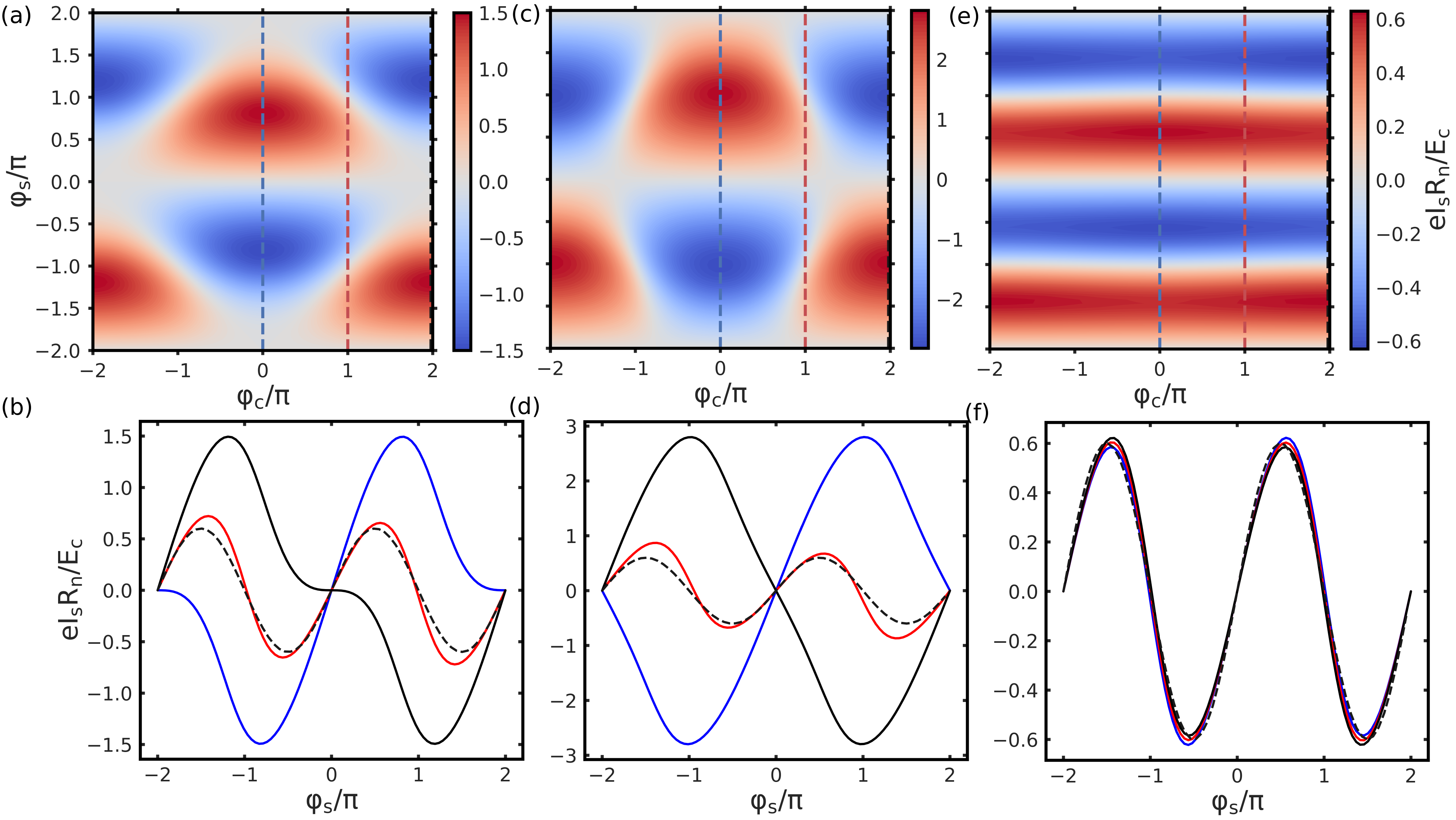}
\caption{ (a)  Normalized supercurrent $I_s$ between the sample superconducting reservoirs as a function of $\varphi_s$ and the phase difference $\varphi_c$ across the other two superconducting reservoirs, with $L_c=L_s=L$.  (b)  $I_s$ as a function of $\varphi_s$ for three different values of $\varphi_c$, blue: $\varphi_c=0$; red: $\varphi_c=\pi$; black: $\varphi_c=2\pi$, corresponding to the cuts shown in (a).  Dashed line corresponds to a pure sine function. (c)  Similar to (a), except with $L_c=0.5L$ and $L_s=L$. (d)  Similar curves as in (b), but corresponding to (c).  (e)  $I_s$ as a function of $\varphi_c$ and $\varphi_s$ with $L_c=4L$, $L_s=L$.  (f)  Similar to (b) and (d), but corresponding to (e). }
\label{fig:Fig2}
\end{figure*}

The simplest geometry in which these effects can be seen is shown schematically in Fig. \ref{fig:Fig1}.  It consists of four normal metal wires connected to each other on one end, and connected to 4 separate superconducting reservoirs on their other ends.  Phase differences $\varphi_s$ and $\varphi_c$ are dropped symmetrically across pairs of superconducting reservoirs, with the corresponding lengths of the normal wires from the reservoirs to the central node being $L_s$ and $L_c$, as shown in the schematic.  To calculate the supercurrents generated in response to these phases, we shall use the quasiclassical equations of superconductivity in the diffusive limit.  These have been discussed in detail elsewhere \cite{rammer_quantum_1986,belzig_quasiclassical_1999}, but the main equations and details of the simulation are reviewed in the Appendix.  The fundamental energy scale of the proximity effect in the diffusive limit is the Thouless energy $E_c= \hbar D/L^2$, where $D=(1/3) v_F \ell$ is the electronic diffusion coefficient, $v_F$ being the Fermi velocity and $\ell$ the elastic mean free path of the electrons in the normal metal.  $L$ is a representative length of the normal metal that will be specified later.  The other relevant energy scale is the superconducting gap $\Delta$, which we take to be $\sim 15 E_c$ for these simulations, based on experiments \cite{NohNonlocal}. In what follows, we shall denote the pair of superconducting reservoirs with the phase difference $\varphi_s$ between them as defining the `sample' junction, while the pair of superconducting reservoirs with phase difference $\varphi_c$ form the `control' junction.  

\begin{figure*}
\includegraphics[width=17.5cm]{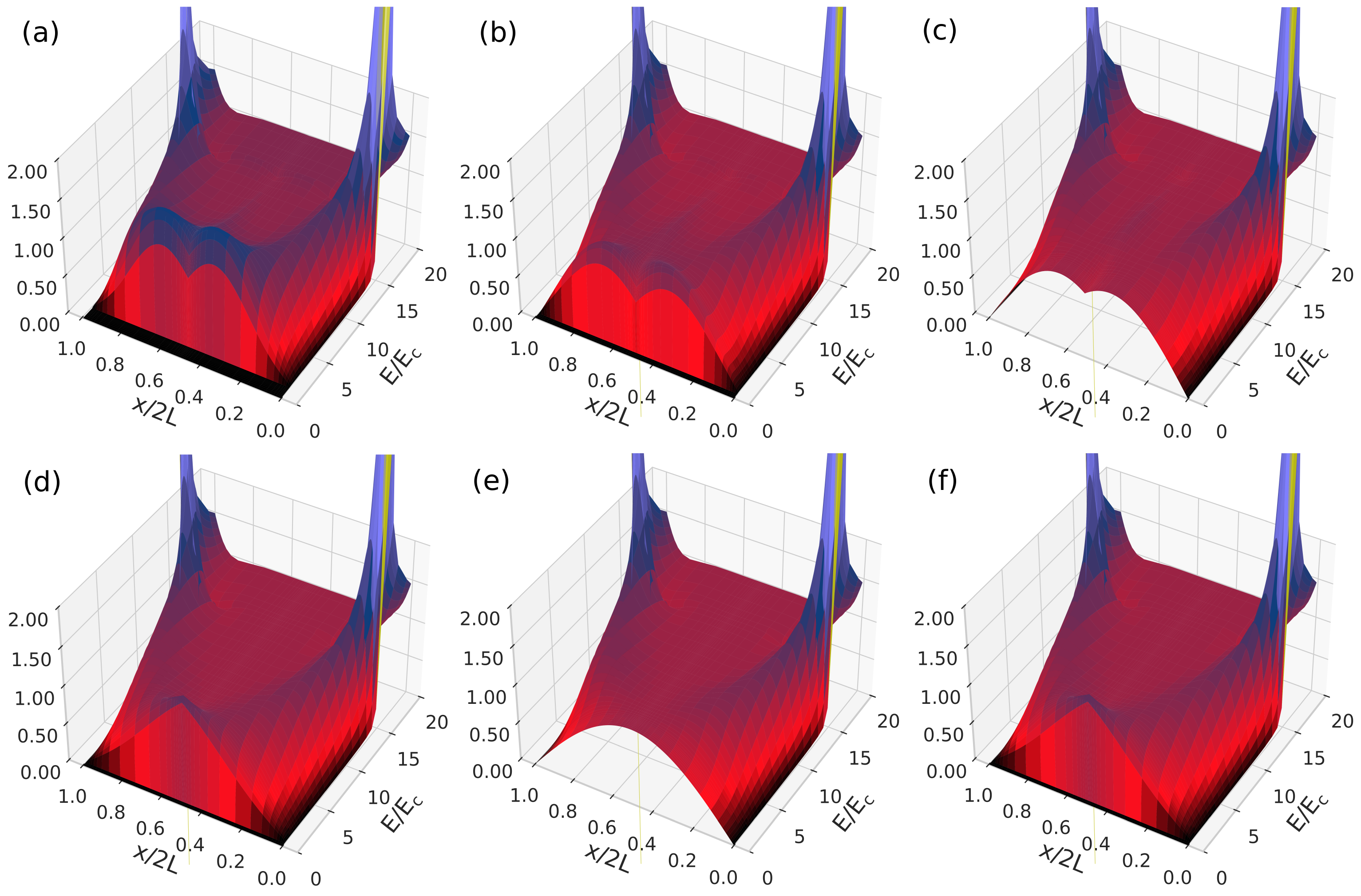}
\caption{ Normalized quasiparticle density of states $N_s(E)/N(0)$ as a function of energy $E$ along the length of the sample junction (between the superconducting reservoirs with phase difference $\varphi_s$) for various values of $\varphi_c$ and $\varphi_s$.  $L_c=0.5L$ and $L_s=L$ (see schematic in the inset to Fig. \ref{fig:Fig1}(a).)  (a) $\varphi_c=0$, $\varphi_s=0$; (b) $\varphi_c=0$, $\varphi_s=\pi$; (c) $\varphi_c=0$, $\varphi_s=2\pi$; (d) $\varphi_c=\pi$, $\varphi_s=0$; (e) $\varphi_c=\pi$, $\varphi_s=\pi$; (a) $\varphi_c=\pi$, $\varphi_s=2\pi$.}
\label{fig:Fig3}
\end{figure*}

Figure \ref{fig:Fig2}(a) shows the $I_s$ between the sample superconducting reservoirs as a function of the phase differences $\varphi_c$ and $\varphi_s$, normalized to $E_c/eR_N$, where $R_N$ is the normal-state resistance of the junction.  Here the length of all the normal wires ($L_s,L_c$) is unity in units of $L$.  While obvious from the plot, the first thing to note that $I_s$ is a function of both $\varphi_c$ and $\varphi_s$.  What this means is that the current-phase relation of the sample junction is a function of control junction phase difference $\varphi_c$.  In particular, the maximum magnitude of $I_s$ for any value of $\varphi_s$, i.e., the critical current $I_c$ of the second junction depends on $\varphi_c$, being a maximum for $\varphi_c=0$ and a minimum for $\varphi_c \sim \pi$.  Thus one can control the critical current of one junction in this multiterminal Josephson junction by modulating the phase across the second junction, the maximum variation here being approximately a factor of 2.      

More surprising, however, is that while $I_s$ is a periodic function of $\varphi_s$, the fundamental period is not $2\pi$ as would be expected from a simple SNS junction, but $4\pi$.  This can be seen more clearly if we plot $I_s$ as a function of $\varphi_s$ for a few select values of $\varphi_c$ (0, $\pi$ and $2\pi$), as shown in Fig. \ref{fig:Fig2}(b).  For $\varphi_c=0$ or $2\pi$ the fundamental period of the current phase relation is clearly $4\pi$, although the functional form is highly non-sinusoidal.  For $\varphi_c=\pi$, the current-phase relation appears sinusoidal with a period of $2\pi$.  Comparison with a pure sine function (shown as the dashed line in Fig. \ref{fig:Fig2}(b)), however, shows that it is not strictly sinusoidal, but a very good approximation.  If we change the length of the normal wires, still keeping all the lengths the same, the overall shape of the $I_s-\varphi_s$ curves and their dependence on $\varphi_c$ does not change, but the magnitude of $I_s$ decreases almost exponentially with increasing length (not shown), as might be expected from earlier experiments and theory \cite{Dubos}.

We now shorten the length of the control junction ($L_c=L/2$), keeping the length of the sample junction the same ($L_s=L$).  The resulting $I_s$ as a function of $\varphi_c$ and $\varphi_s$ is shown in Fig. \ref{fig:Fig2}(c).  While the length $L_S$ is the same as in Fig. \ref{fig:Fig2}(a), the magnitude of $I_s$ has increased approximately twofold.   The shape of the $I_s-\varphi_s$ curve has also changed:  at $\varphi_c=0$ and $2\pi$, it now more closely approximates a sine function, but with a fundamental period of $4\pi$.  When $\varphi_c = \pi$,  $I_s-\varphi_s$ again approximates a sine function, but now with a period of $2\pi$.  Note also that the variation in the critical current with $\varphi_c$ is larger, roughly a factor of 3.  Figures \ref{fig:Fig2}(a)-(d) demonstrate that by modulating the phase across one pair of superconducting reservoirs, one can tune significantly not only the magnitude of the critical current between another pair of superconducting reservoirs, but also the fundamental periodicity of the current-phase relation.

The modulation of both the amplitude and the current-phase relation depends critically on the length of the control junction $L_c$, being stronger when this length is shorter.  To demonstrate this, Fig. \ref{fig:Fig2}(e) shows $I_s$ as a function of $\varphi_c$ and $\varphi_s$ with $L_s=L$ as before, but now with $L_c=4L$.  The result is strikingly different from Figs. \ref{fig:Fig2}(a) and (c):  there is only a weak dependence of $I_s$ on $\varphi_c$.  Cuts taken at $\varphi_c=0$, $\pi$, and $2\pi$ as in Figs.\ref{fig:Fig2}(b) and (d) confirm this, showing three almost identical curves that closely approximate a sine function with a fundamental period of $2\pi$, as would be expected for a single SNS junction, with a reduced $I_c$ in comparison the the previous two cases, even though $L_s$ is the same length.

Further insight can be gained by studying the quasiparticle density of states $N_s(E)$ along the length of the sample junction at various values of $\varphi_c$ and $\varphi_s$.  Before doing this, let us review what is expected for a single SNS junction as a function of the phase difference $\varphi$ across it \cite{minigap}.  For $\varphi=0$ a minigap of order $\approx E_c$ is present in the normal metal such that at energies $E\lessapprox E_c$, $N_s(E)=0$.  Since $\varphi=0$, the supercurrent $I_s$ also vanishes.  As $\varphi$ increases, $I_s$ first increases, reaching a maximum at some value of $\varphi$ between 0 and $\pi$ (at $\pi/2$ if the current-phase relation is sinusoidal).  As $\varphi$ is increased in this range, the minigap progressively decreases in magnitude, so that eventually at $\varphi=\pi$,  both the minigap and the supercurrent $I_s$ vanish.  As $\varphi$ is increased further, both the minigap and supercurrent reappear such that at $\varphi=2\pi$ we are back again to the full minigap with $I_s=0$, similar to the situation with $\varphi=0$.  This corresponds to a fundamental periodicity of the current-phase relation of $2\pi$.

With this behavior of $N_s(E)$ for a single SNS junction in mind, let us now look at $N_s(E)$ in the 4 terminal junction.  Figure \ref{fig:Fig3} shows $N_s(E)$ normalized to the normal density of states $N(0)$ along the length of the sample junction as a function of energy $E$.  Here the dimensions of the device are the same as in Figs. \ref{fig:Fig2}(c) and (d) ($L_c=0.5 L$, $L_s=L$) such that the midpoint of the wire (i.e, the node) is at $x=L$.  Consequently, we plot $N_s(E)$ as a function of $x/2L$ and $E$.  Since we have assumed transparent interfaces, $N_s(E)$ at $x=0$ and $x=2L$ (the superconducting reservoirs) corresponds to the density of states for a superconductor with the gap $\Delta \sim 15 E_c$, as seen in all the panels in Fig. \ref{fig:Fig3}.  

Figures \ref{fig:Fig3}(a), (b) and (c) show $N_s(E)/N(0)$ for $\varphi_s=0$, $\varphi_s=\pi$ and $\varphi_s=2\pi$, with $\varphi_c=0$.  For $\varphi_s=0$, there is a small uniform minigap of order $\sim 1.5 E_c$ along the entire length, although $N_s(E)$ at higher energies is not uniform.  For $\varphi_s=\pi$, there is still a uniform minigap, but now with reduced magnitude of $\sim 0.5 E_c$.  Finally, for $\varphi_s=2\pi$, there is no gap in the density of states along the wire (except at the reservoirs).  In this sense  $N_s(E)$ for $\varphi_s=2\pi$ is similar to what is expected for a simple SNS junction for $\varphi=\pi$, an indication that the fundamental periodicity in $\varphi_s$ is $4\pi$.

Figures \ref{fig:Fig3}(d), (e) and (f) show similar data for the case of $\varphi_c=\pi$.  For $\varphi_s=0$, we again have a uniform minigap, albeit with a lower magnitude of $\approx  0.5 E_c$ compared to the $\varphi_c=0$ case.  For $\varphi_s=\pi$, however, the minigap vanishes, while for $\varphi_s=2\pi$ $N_s(E)$ is similar to that at $\varphi_s=0$, with a minigap of the same magnitude.  This is in fact what we would expect to see for a single SNS junction whose current-phase relation is $2\pi$ periodic.

The tuning of the current-phase relation of the sample junction by the phase difference applied across the control junction can be understood using a simple circuit model of the supercurrents in the device.  Consider again the device schematic shown in Fig. \ref{fig:Fig1}.  The red and green arrows show two possible paths that the supercurrent can take between the superconducting reservoir with phase $\varphi_s/2$ to the reservoir with phase $-\varphi_s/2$.  One is directly along the normal wires joining the two reservoirs (red line), and the other is via the control superconducting reservoir with phase $-\varphi_c/2$ (green lines).  (There will also be a similar green pathway through the control reservoir with phase $\varphi_c/2$.)  As with a simple SNS junction, we assume that the supercurrent between any two superconducting reservoirs depends on the phase difference between them and the corresponding current-phase relation is $2\pi$ periodic, and that the magnitude of the supercurrent between two superconducting reservoirs is a strongly decaying function of the length of the normal metal between them.  This is borne out by previous experimental and theoretical work \cite{Dubos} and our simulations as noted earlier. 
For simplicity, we also assume that the current-phase relation is sinusoidal, although that is not necessary for the argument. 

Let us first consider the case when $\varphi_c=0$ and $\varphi_s=\pi$.  The (red) supercurrent between the reservoirs with phase $\pm \varphi_s/2$ vanishes as their phase difference is $\pi$.  However, the phase difference between the sample junction superconducting reservoirs and the control junction reservoirs is $\pm \pi/2$, and hence the (green) supercurrents between them have their maximum value.  On the other hand, if $\varphi_c=\pi$ while still keeping $\varphi_s=\pi$, then the phase difference between any pair of superconducting reservoirs is either 0 or $\pi$, hence no supercurrent flows through the device.

In general, the net supercurrent between the sample superconducting reservoirs is the sum of the direct (red) supercurrent between them and the (green) supercurrents via the control superconducting reservoirs, the former being $2\pi$ periodic while the latter are $4\pi$ periodic with respect to $\varphi_s$.  If the lengths $L_c$ and $L_s$ are the same, then the contribution of these currents is nominally equal, but their relative contributions can be tuned by varying $\varphi_c$, as seen in Figs. \ref{fig:Fig2}(a) and (b).  If $L_c$ is shorter than $L_s$, then the $4\pi$ periodic currents will dominate, but can be turned off by setting $\varphi_c=\pi$.  Finally, if $L_c$ is longer than $L_s$, then the $2\pi$ periodic supercurrent dominates, and we lose the ability to tune the current-phase relation by varying $\varphi_c$.  As the magnitude of the supercurrent is expected to vary exponentially with distance between superconducting reservoirs at finite temperatures \cite{Dubos}, this is expected to be a strong effect.  

In summary, a 4-terminal diffusive Josephson junction thought of as two individual Josephson junctions allows the ability to tune both the magnitude of the critical current and the current-phase relation of one of the junctions by tuning the phase difference across the other junction.  In particular, with suitable device geometry, this allows changing the fundamental periodicity of the current-phase relation from $2\pi$ to $4\pi$.  The effect should be observable by measuring the density of states at the central node through a tunneling measurement in a device where the phases $\varphi_c$ and $\varphi_s$ can be statically and independently by connecting the superconductors to form flux loops, as has been experimentally demonstrated recently for devices with 3 superconducting contacts \cite{coraiola_phase-engineering_2023, strambini_-squipt_2016, wisne_topo}.  It could also be observed in devices with additional normal contacts to the central normal metal, as was already done for 3 terminal devices \cite{wisne_topo}.  Here the oscillations of the resistance as a function of the flux coupled to the flux loops should reveal the periodicity of the current-phase relation, although the detailed structure might be modified by the addition of the normal probes.  The ability to tune the current-phase relation may have applications in controlling the behavior of superconducting qubits, where the integral of the current-phase relation determines the potential term that goes into the qubit Hamiltonian.

\begin{acknowledgments}
I thank K. Ryan and M. Wisne for a critical reading of the manuscript. 
 This research was conducted with support from the National Science Foundation under Grant No. DMR-2303536.
\end{acknowledgments}

\vspace{1cm}
\appendix
\noindent \textbf{Appendix}\\
The proximity effect in the normal wires in the geometry of Fig. \ref{fig:Fig1}(a) is encapsulated by the energy and space dependent quasiclassical Green's function $\hat{g}_s$ which satisfies the Usadel equation \cite{Chandrasekhar2008}
\begin{equation}
    [\hat{\tau}^3 E + \hat{\Delta}, \hat{g}_s] = i D \partial_{\vec{R}}(\hat{g}_s \partial_{\vec{R}} \hat{g}_s).
    \label{eq:usadel}
\end{equation}
Here $E$ is the energy, $\partial_{\vec{R}}$ is the spatial gradient and $D=(1/3) v_F \ell$ is the electronic diffusion coefficient, $v_F$ being the Fermi velocity and $\ell$ the elastic mean free path.  $\hat{\Delta}$ and $\hat{\tau}^3$ are 4x4 matrices that represent the gap and the Pauli matrix in the combined Keldysh/Nambu-Gorkov space
\begin{equation}
    \hat{\Delta} =
    \begin{pmatrix}
        \tilde{\Delta} & 0 \\
        0  & \tilde{\Delta}
    \end{pmatrix},
    \quad \hat{\tau}^3 =
        \begin{pmatrix}
        \tau^3 & 0 \\
        0  & \tau^3
    \end{pmatrix}    
\end{equation}
with
\begin{equation}
    \tilde{\Delta} =
    \begin{pmatrix}
        0 & \Delta \\
        - \Delta ^*  & 0
    \end{pmatrix},
    \quad \tau^3 =
        \begin{pmatrix}
        1 & 0 \\
        0  & -1
    \end{pmatrix}    
\end{equation}
and $\Delta$ is the complex order parameter.  $\hat{g}_s$ is also represented as a 4x4 matrix
\begin{equation}
  \hat{g}_s =
  \begin{pmatrix}
      \tilde{g}_s^R & \tilde{g}_s^K \\
      0    & \tilde{g}_s^A
  \end{pmatrix}
\end{equation}
where the 2x2 matrices $\tilde{g}_s^R$, $\tilde{g}_s^A$ and $\tilde{g}_s^K$ are the retarded, advanced and Keldysh components of the quasiclassical Green's function in Nambu-Gorkov space.  The Keldysh component of the Green's function $\tilde{g}_s^K$ can be written in terms of $\tilde{g}_s^R$, $\tilde{g}_s^A$ and the quasiparticle distribution function $\tilde{h}$ as 
\begin{equation}
   \tilde{g}_s^K =  \tilde{g}_s^R \tilde{h} - \tilde{h} \tilde{g}_s^A.
\end{equation}
The diagonal components of the Usadel equation, Eqn. \ref{eq:usadel} involve only $\tilde{g}_s^R$ or $\tilde{g}_s^A$ and hence the equilibrium properties of the system, while the (12) component of the equation involves $\tilde{g}_s^K$ and hence $\tilde{h}$, and therefore describes the nonequilibrium properties of the system. $\tilde{h}$ is usually represented in a form diagonal in Nambu-Gorkov space as
\begin{equation}
    \tilde{h} = h_T \tau^0 + h_L \tau^3
\end{equation}
where $\tau^0$ is the 2x2 identity matrix, and $h_T$ and $h_L$ are the so-called transverse and longitudinal distribution function introduced by Schmid and Sch\"on \cite{schmid_linearized_1975} whose equilibrium forms will be defined later.

The charge current in the system can be represented in the form
\begin{equation}
    \vec{j}(\vec{R}, T) = e N_0 D \int dE (M_{33}(\partial_{\vec{R}} h_T) + Q h_L + M_{03} (\partial_{\vec{R}} h_L)) 
\label{eq:chargecurrent}
\end{equation}
where $N_0$ is the normal density of states at the Fermi energy, and $Q$ and $M_{ij}$ are the spectral supercurrent and normalized energy and space dependent diffusion coefficients given by
\begin{align}
    Q &= \frac{1}{4} \mathtt{Tr}\{\tau^3[\tilde{g}_s^R(\partial_{\vec{R}}\tilde{g}_s^R) - \tilde{g}_s^A(\partial_{\vec{R}}\tilde{g}_s^A)]\} \\
    M_{ij} &= \frac{1}{4} \mathtt{Tr} \{\delta_{ij} \tau^0 - \tilde{g}_s^R \tau^i \tilde{g}_s^A \tau^j\} 
\end{align}
where the $\tau^i$ are the Pauli matrices.  

The Usadel equation in the proximity-coupled normal wires needs to be solved subject to boundary conditions specified at the reservoirs and at the node.  In order to facilitate the numerical solution of the equation, one can use the representation of the Green's function in terms of the complex Riccati parameters $\gamma$ and $\tilde{\gamma}$ \cite{NohNonlocal}
\begin{equation}
 \tilde{g}_s^R   = \frac{1}{1+\gamma \tilde{\gamma}}
 \begin{pmatrix}
     1 - \gamma \tilde{\gamma} & 2 \gamma \\
     2 \tilde{\gamma} & \gamma \tilde{\gamma} - 1
 \end{pmatrix}
\end{equation}
with $\tilde{g}_s^A = - \tau^3 (\tilde{g}_s^R)^\dagger \tau^3$.  This representation satisfies the quasiclassical Green's function normalization condition $(\tilde{g}_s^R)^2 = \tau^0$.  At the node where all the normal metal wires meet in Fig. 1, continuity of the Green's function gives the Kirchhoff like boundary condition $ \sum_i \tilde{g}_{si}^R (\partial_{\vec{R}}\tilde{g}_{si}^R)=0$
where the sum is over all the normal metal wires connecting at the node, with a similar condition for the charge currents (Eqn. \ref{eq:chargecurrent}) entering the node.  The boundary condition at the interface between a normal metal wire and a superconducting reservoir are determined by the transparency of the interface between them.  Here I assume that these interfaces are perfectly transparent, so that the value of the Green's function in the normal wire at these interfaces is given by the equilibrium value of the Green's functions in the reservoirs.  In terms of the Riccati parameters, these equilibrium values are given by
\begin{align}
  \gamma_0(E) &= \frac{\Delta_0}{E + \sqrt{E^2 - |\Delta_0|^2}}  \\
  \tilde{\gamma}_0(E) &= -\frac{\Delta_0^*}{E + \sqrt{E^2 - |\Delta_0|^2}}.
\end{align}
where $E$ is the energy.  The phase of the superconductor can be taken into account through the phase of the order parameter:  $\Delta_0 \rightarrow \Delta_0 e^{i\varphi}$.  Finally, $h_T$ and $h_L$ are given by their equilibrium values
\begin{equation}
    h_{L,T} = \frac{1}{2} \left[ \tanh \left(\frac{E+eV}{2 k_B T}\right) \pm \tanh \left(\frac{E-eV}{2 k_B T}\right) \right]
\end{equation}
where $T$ is the temperature and $V$ is the voltage on the reservoir.
In this case, $V=0$ on the superconducting reservoirs so that $\partial_{\vec{R}} h_T=\partial_{\vec{R}} h_L=0$ and the only term contributing to the charge current in Eq. (\ref{eq:chargecurrent})is the one involving the spectral supercurrent $Q$.  For geometries like that in Fig. 1, where the normal wires are effectively one-dimensional, the Usadel equation can be normalized to a length $L$, $x\rightarrow x/L$, where $x$ is the position along the wire.  This naturally leads to the characteristic energy scale $E_c = \hbar D/L^2$ of the problem, the Thouless energy, in which the other energy scales of the problem can be expressed.  For the simulations presented here, based on experimentally relevant parameters \cite{NohNonlocal}, the superconducting gap is chosen to be $\Delta \approx  15 E_c$ and the temperature $T\approx 0.2 E_c$, corresponding to the low temperature limit.  The simulations were performed using the open-source Usadel solver of Pauli Virtanen \cite{virtanen_thermoelectric_2007}.

%

\end{document}